\documentclass[pre,twocolumn,showpacs]{revtex4}
\usepackage{graphicx}
\usepackage{color}
\usepackage{amsmath}

\begin{document}

\title{Heat devices in nonlinear irreversible thermodynamics}

\author{Y. Izumida}
\thanks{Present address: Department of Complex Systems Science, Graduate School of Information Science, Nagoya University, Nagoya 464-8601, Japan}
\affiliation{Department of Information Sciences, Ochanomizu University 2-1-1 Otsuka, Bunkyo-ku, Tokyo 112-8620, Japan}
\author{K. Okuda}
\affiliation{Division of Physics, Hokkaido University, Sapporo 060-0810, Japan}
\author{J. M. M. Roco and A. Calvo Hern\'andez }
\affiliation{Departamento de F{\'\i}sica Aplicada, and
Instituto Universitario de F\'{\i}sica Fundamental y Matem\'aticas (IUFFyM),
Universidad de Salamanca, 37008 Salamanca, Spain}
\date{\today}

\begin{abstract}
We present results obtained by using nonlinear irreversible models for heat devices. In particular, we focus on the global performance characteristics, the maximum efficiency and the efficiency at maximum power regimes for heat engines, and the maximum coefficient of performance (COP) and the COP at maximum cooling power regimes for refrigerators. We analyze the key role played by the interplay between irreversibilities coming from heat leaks and internal dissipations. 
We also discuss the relationship between these results and those obtained by different models.
\end{abstract}

\maketitle

\section{Introduction}\label{s1}

Nowadays the thermodynamic description of the efficient performance regimes in heat devices has
been becoming of special relevance, due to the growing importance
of saving energy resources in any operation of energy conversion. The Carnot's theorem states
the upper bounds of heat-energy conversion processes between two heat reservoirs at
temperatures $T_{c}$ and $T_{h}$ ($T_{c}<T_{h})$: for a heat device working as a heat engine (HE) the maximum
efficiency is $\eta_{\mathrm{C}}=1-\tau$ ($\tau \equiv T_{c}/T_{h}$), while working as a refrigerator (RE) the maximum coefficient of performance (COP) is $\varepsilon_{\mathrm{C}}=\tau/(1-\tau)$. However, 
these upper bounds are of no practical use since they
refer to reversible cycles with zero power output for HE and zero cooling power for RE. For the
thermodynamic analysis of real heat devices (working at non-zero
rate along irreversible paths), different models and different
figures of merit (based on thermodynamic, economic, compromised, and
sustainable considerations) have been proposed~\cite{bejan96, lchen99,berry00,wu04,durmayaz04}.

A part of such models is founded on finite-time thermodynamics (FTT) considerations. FTT focuses on
irreversibilities caused by finite-rate heat transfers
between the working fluid, and the external heat reservoirs,
internal dissipation of the working fluid and heat leaks between the heat
reservoirs. The optimization procedure in FTT, carried out under a fixed cycle time, 
usually assumes two degrees of freedom, that is, the inner temperatures of
the isothermal steps of the working system~\cite{lchen99,wu04,durmayaz04,jchen01,luisarias13}.
In spite of their analytical simplicity, these models can reproduce, at least qualitatively,
the power-efficiency and the cooling power-COP behaviors observed in real HE~\cite{jchen01} and
RE~\cite{gordon00,chen97}. More simplified FTT models assume the so-called
endoreversible approximation~\cite{lchen99,wu04,jchen01,luisarias13,gordon00}, 
where the heat leaks and the internal dissipations are neglected. In this case,
if the linear heat transfer laws are additionally assumed for the external heat exchange, the resulting efficiency at maximum power becomes the well-known Curzon-Ahlborn (CA) efficiency $\eta_{\mathrm{CA}}=1-\sqrt{\tau}=1-\sqrt{1-\eta_{\rm
C}}$~\cite{curzon75}.

In the recently proposed low-dissipation (LD) model~\cite{esposito10a}, the basic starting
point is that the entropy production in the isothermal hot
(cold) heat exchange process is assumed to behave as $\Sigma_h/t_h$
($\Sigma_c/t_c$), with $t_h$ and $t_c$ denoting the corresponding
time durations, and $\Sigma_h$ and $\Sigma_c$ being coefficients
containing information on the irreversibility sources.
Considering $t_h$ and $t_c$ as degrees of freedom for optimization, the LD model for HE provides the
upper ($\eta_{\rm {maxP}}^+=\frac{\eta_{\rm C}}{2-\eta_{\rm C}}$)
and lower ($\eta_{\rm {maxP}}^-=\frac{\eta_{\rm C}}{2}$) bounds for
the efficiency at maximum power $\eta_{\rm maxP}$ under extremely asymmetric
dissipation limits $\Sigma_h/\Sigma_c\rightarrow \infty$
and $\Sigma_h/\Sigma_c \rightarrow 0$, respectively.
Indeed, the LD model allows us to recover the Curzon-Ahlborn value $\eta_{\rm CA}$ when
symmetric dissipation is considered ($\Sigma_h=\Sigma_c$), in this case without assuming any specific heat transfer law. 
An important result of the LD model for HE is the linking of the
CA efficiency to symmetry considerations, providing a unified
framework to understand the quasi-universal behavior shown by the
efficiency at maximum power of many different kinds of heat engines.
An extension of the LD model to RE was also reported~\cite{tu12d,carla12a,hu13},
but in this case considering, as a figure of merit, the $\chi$ function defined 
as $\chi=\varepsilon \dot Q_c$, being $\varepsilon$ and $\dot Q_c$ the COP and the cooling power, respectively. The lower
($\varepsilon_{{\rm max} \chi}^-=0$) and upper ($\varepsilon_{{\rm max}
\chi}^+=(\sqrt{9+8\varepsilon_{\rm C}}-3)/2$) bounds of the COP at maximum
$\chi$ were obtained under extremely asymmetric dissipation limits
$\Sigma_h/\Sigma_c \rightarrow \infty$ and $\Sigma_h/\Sigma_c \rightarrow 0$, respectively~\cite{tu12d}. Under the symmetric condition ($\Sigma_h=\Sigma_c$), the optimized COP was found to be $\varepsilon_{{\rm
max\chi}}=\sqrt{1+\varepsilon_{\rm C}}-1\equiv \varepsilon_{\rm CA}$~\cite{carla12a}.
This result, obtained previously in different contexts~\cite{yan90,velasco97a,allahv10a},
could be viewed as a counterpart of the CA efficiency for HE, though this point is a current issue of discussion~\cite{apertet13a} (see Sec.~V).

Linear irreversible thermodynamics (LIT) is a well-founded formalism, which is focused on the irreversible evolution of macroscopic
systems allowing us to extend the scope of the equilibrium thermodynamics. LIT assumes systems in local equilibrium and defines thermodynamic
forces and fluxes both interlinked by means of linear relationships governing the macroscopic evolution. From the
LIT principles, it is possible to construct models of heat devices~\cite{vbroeck05,yuki09,arias08,apertet13b,cisneros06,cisneros08,tu13b,yuki14a}, from which 
the optimization procedure considers the thermodynamic force as the sole relevant degree of freedom. In LIT the endoreversible features of the simplified FTT models (including the CA model~\cite{curzon75}) of heat devices are recovered under the so-called tight-coupling condition~\cite{vbroeck05,cisneros06,cisneros08}.

Minimally nonlinear irreversible thermodynamic (MNLIT) models have also
been proposed for cyclic and steady-state heat devices, in order
to account for possible thermal dissipation effects in the interaction between the working
system and the external heat reservoirs~\cite{yuki12a,yuki13a}. 
These models incorporate the Onsager relations with an additional nonlinear dissipation term.
As in LIT, the optimization procedure also involves only one degree of freedom (the thermodynamic flux), 
and the outstanding results provided by the low-dissipation models~\cite{esposito10a,tu12d,carla12a} are recovered by the MNLIT model as a particular case~\cite{yuki12a} and~\cite{yuki13a} (see Appendix A). 
We stress that a numerical validation of the LD-model and thus the MNLIT model has been confirmed in a computer simulation of a Carnot cycle with a single particle~\cite{hoppenau13a}.
However, the MNLIT models are subject to some criticism. In Ref.~\cite{apertet13b} it is claimed that the MNLIT models are misleading, since dissipations should naturally appear in the LIT models when the local Onsager relations are extended to a global scale. Indeed, for the thermoelectric devices these authors of Ref.~\cite{apertet13b} showed that the Joule dissipation is well-founded in LIT based on this argument. However, for any generic heat device, the relation between the local and global scales may be too complicated or it simply has not been obtained yet. In such cases, the addition of the dissipative terms adopted in the MNLIT models could be considered as a reasonable and natural assumption in order to explain the thermodynamic (macroscopic) influence of the local dissipative effects on a generic heat device.

The present paper is aimed to present results obtained by the MNLIT models for both HE and RE, and it has two main goals: (i) to present a unified description of non-isothermal heat devices for HE and RE, making emphasis on the global performance characteristics of the behaviors of power-efficiency and cooling power-COP curves, including the non-tight-coupling case; and (ii) to analyze the maximum efficiency and the efficiency at maximum power regimes for HE and the maximum COP and the COP at maximum cooling power regimes for RE. Especially, we reveal the impacts of irreversibilities by heat leaks (degrees of the coupling strength), internal dissipations, and their interplay on the performance of the heat devices, by considering their various limits. 

To attain these goals, after giving a brief theoretical background in Sec.~II, we present in Sec.~III
a detailed analysis of the main global performance characteristics of HE and RE in terms of the degrees of the coupling strength and the dissipation effects. Then Sec.~IV focuses on the optimum performance regimes, and, finally, we discuss our main results in Sec.~V.
The original idea of the MNLIT model, which was proposed with the motivation to explain and extend the LD model~\cite{yuki12a}, is also given in Appendix A.

\section{Theoretical background}\label{s2}
Although the main theoretical aspects for both HE and RE have been reported previously~\cite{yuki12a,yuki13a}, here for the sake of completeness, 
we give a brief theoretical background emphasizing the unified framework for HE and RE.
\begin{figure}[h]
\includegraphics[scale=0.3]{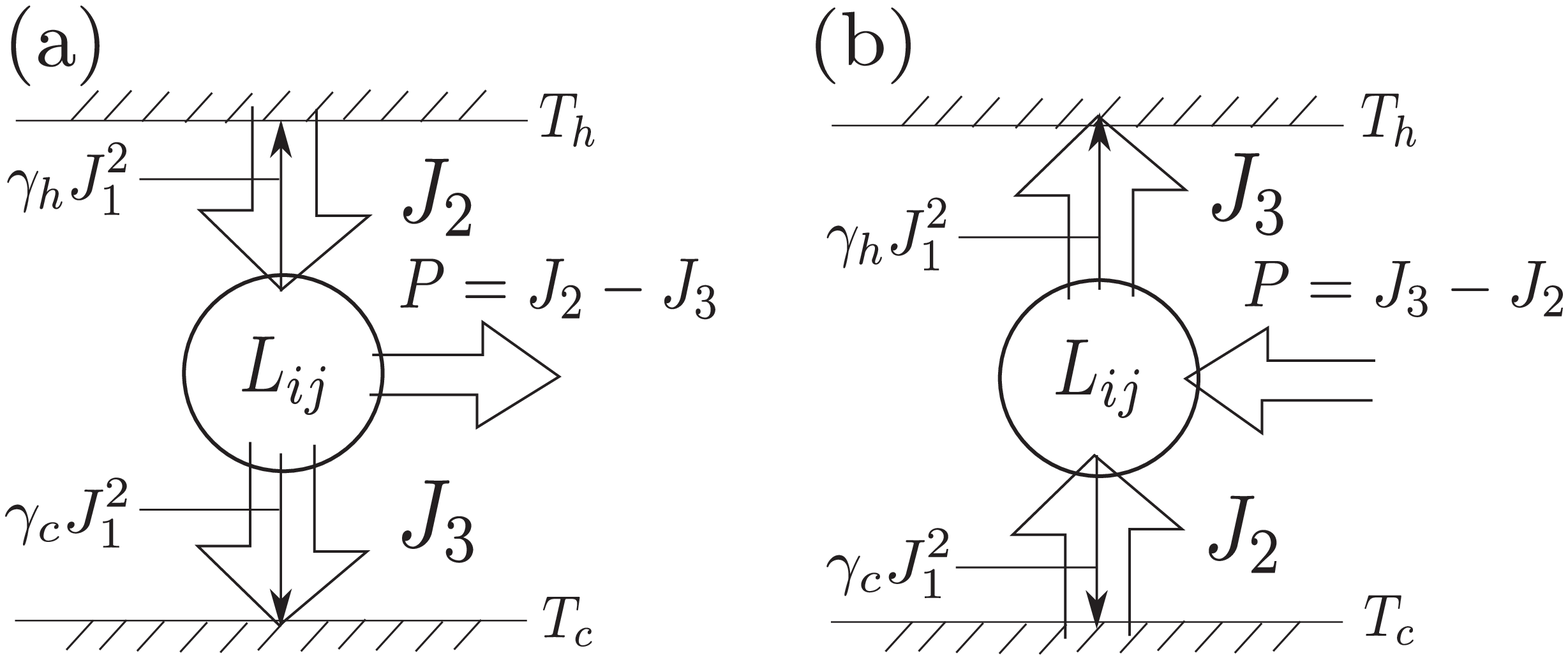}
\caption{Set up of minimally nonlinear irreversible heat devices: (a) heat engine and (b) refrigerator.
The thin arrows inside the bold arrows show the direction of the dissipation terms included in the heat fluxes.
}\label{heat_device}
\end{figure}

\subsection{Minimally nonlinear irreversible model for HE}\label{s2a}
We start from the entropy production rate $\dot{\sigma}$ of the total system (that is, the heat engine and the heat reservoirs).
Hereafter the dot denotes a quantity divided by cycle period for cyclic heat engines or a quantity per unit time for steady-state heat engines. 
Because the internal state of the heat engine comes back to the original state after one cycle for cyclic heat engines 
or it remains unchanged for steady-state heat engines, 
$\dot{\sigma}$ can be written by the sum of the entropy-change rate of the heat reservoirs as
$\dot{\sigma}=\dot{S}_h+\dot{S}_c$ according to the local equilibrium hypothesis~\cite{callen}, where $S_i$ ($i=h, c$) denotes the equilibrium entropy of the heat reservoir.
It can be written as
\begin{eqnarray}
\dot{\sigma}=\dot{S}_h+\dot{S}_c=\dot{U}_h\frac{\partial S_h}{\partial U_h}+\dot{U}_c\frac{\partial S_c}{\partial U_c}=-\frac{\dot{Q}_h}{T_h}+\frac{\dot{Q}_c}{T_c},\label{eq.sigma_dot_he_1}
\end{eqnarray}
where we have used $\frac{\partial S_i}{\partial U_i}=\frac{1}{T_i}$ with $U_i$ being the equilibrium internal-energy of the heat reservoir, and 
for HE, we denote by $\dot{Q}_h\equiv -\dot{U}_h$ the heat flux from the hot heat reservoir and by $\dot{Q}_c\equiv \dot{U}_c$ the heat flux into the cold heat reservoir, respectively.
We also denote by $\dot{W}\equiv P$ the power output.
Then, from the relation $\dot{Q}_c=\dot{Q}_h-P=\dot{Q}_h-F\dot{x}$ with $F$ and $x$ a generalized external force and its conjugate variable, respectively,  
Eq.~(\ref{eq.sigma_dot_he_1}) can be rewritten as 
\begin{eqnarray}
\dot{\sigma}=-\frac{F\dot{x}}{T_c}+\dot{Q}_h\left(\frac{1}{T_c}-\frac{1}{T_h}\right)\equiv J_1X_1+J_2X_2.\label{eq.sigma_dot_he_2}
\end{eqnarray}
It naturally leads us to define the thermodynamic flux $J_1\equiv \dot{x}$
(the motion speed of the heat engine)
conjugate to the thermodynamic force $X_1 \equiv -F/T_c$, and the other thermodynamic flux
$J_2\equiv \dot{Q}_h$ (the heat flux from the hot heat reservoir)
conjugate to the other thermodynamic force $X_2\equiv 1/T_c-1/T_h$, where these quantities are expressed by using the thermodynamic extensive and intensive parameters of the equilibrium heat reservoirs~\cite{callen}.
By expanding the thermodynamic flux $J_i$ by the thermodynamic force $X_i$ around the equilibrium point $X_1=X_2=0$ as
$J_i=\sum_{j=1}^2 L_{ij}X_j+\sum_{j,k=1}^2 M_{ijk}X_jX_k+\sum_{j,k,m=1}^2 N_{ijkm}X_jX_kX_m+\cdots$
with $L_{ij}$'s, $M_{ijk}$'s, and $N_{ijkm}$'s being the expansion coefficients of each order, we obtain a full description of the evolution of the entropy production rate of the nonequilibrium heat engine.
The LIT model assumes the following linear Onsager relations between the thermodynamic fluxes and forces~\cite{vbroeck05}:
\begin{eqnarray}
&&J_1=\dot{x}=L_{11}X_1+L_{12}X_2,\label{eq.onsager_J1_he}\\
&&J_2=\dot{Q}_h=L_{21}X_1+L_{22}X_2,\label{eq.onsager_J2_he}
\end{eqnarray}
where the coefficients $L_{ij}$'s are the Onsager coefficients satisfying the reciprocal relation $L_{12}=L_{21}$.
In the LIT model, 
the entropy production rate $\dot{\sigma}=J_1X_1+J_2X_2$ becomes a quadratic form in $X_i$'s and its non-negativity leads to the following restriction on the Onsager coefficients $L_{ij}$'s as
\begin{eqnarray}
L_{11}\ge 0, \ L_{22}\ge 0, \ L_{11}L_{22}-L_{12}^2\ge0.\label{eq.restrict_onsager_coeffi_he}
\end{eqnarray}
By changing the variable from $X_1$ to $J_1$ by using Eq.~(\ref{eq.onsager_J1_he}), we can write
the Onsager relation Eq.~(\ref{eq.onsager_J2_he}) and $\dot{Q}_c=J_2-P\equiv J_3$ by using $J_1$ as
\begin{eqnarray}
&&J_2=\frac{L_{21}}{L_{11}}J_1+L_{22}(1-q^2)X_2,\label{eq.lin_J2_by_J1_he}\\
&&J_3=\frac{L_{21}T_c}{L_{11}T_h}J_1+L_{22}(1-q^2)X_2+\frac{T_c}{L_{11}} {J_1}^2, \label{eq.lin_J3_by_J1_he}
\end{eqnarray}
respectively, where $q$ is the coupling strength parameter defined as
\begin{eqnarray}
q\equiv \frac{L_{12}}{\sqrt{L_{11}L_{22}}} \ \ (|q|\le 1).\label{eq.q_he}
\end{eqnarray}
From Eqs.~(\ref{eq.lin_J2_by_J1_he}) and (\ref{eq.lin_J3_by_J1_he}), we immediately notice that the nonlinear term $\frac{T_c}{L_{11}}J_1^2$ appears only in $J_3$ in an ``asymmetric" way.
By using Eqs.~(\ref{eq.lin_J2_by_J1_he}) and (\ref{eq.lin_J3_by_J1_he}), the entropy production rate $\dot{\sigma}=-\frac{J_2}{T_h}+\frac{J_3}{T_c}$ is given by
\begin{eqnarray}
\dot{\sigma}=L_{22}(1-q^2)X_2^2+\frac{J_1^2}{L_{11}}, 
\end{eqnarray}
where it turns out that the nonlinear term expresses the dissipation effect contributing to the entropy production rate.

Our minimally nonlinear irreversible heat engine assumes the following extended Onsager relations 
such that the dissipation terms in both sides of the heat fluxes are equally taken into account~\cite{yuki12a} [see Fig.~\ref{heat_device} (a)]:
\begin{eqnarray}
&&J_1=\dot{x}=L_{11}X_1+L_{12}X_2,\label{eq.exonsager_J1_he}\\
&&J_2=\dot{Q}_h=L_{21}X_1+L_{22}X_2-\gamma_h J_1^2.\label{eq.exonsager_J2_he}
\end{eqnarray}
The nonlinear term in $J_2$ expresses the dissipation into the hot heat reservoir 
caused by the finite-time motion of the heat engine $J_1\ne 0$ and $\gamma_h$ ($>0$) is a constant meaning the strength of the dissipation.
This choice of the specific form is motivated by the low-dissipation Carnot cycle model, which is proved to be equivalent with
Eqs.~(\ref{eq.exonsager_J1_he}) and (\ref{eq.exonsager_J2_he}) under the tight-coupling condition~\cite{yuki12a}
(see Appendix A). In the absence of the nonlinear term, Eqs.~(\ref{eq.exonsager_J1_he}) and (\ref{eq.exonsager_J2_he}) recover the usual linear Onsager relations Eqs.~(\ref{eq.onsager_J1_he}) and (\ref{eq.onsager_J2_he}). 
Although our extended Onsager relation Eq.~(\ref{eq.exonsager_J2_he}) includes the additional nonlinear term,
we assume that the Onsager reciprocity $L_{12}=L_{21}$ and the restriction Eq.~(\ref{eq.restrict_onsager_coeffi_he}) still holds for our $L_{ij}$'s (Appendix A).

By changing the variable from $X_1$ to $J_1$ by using Eq.~(\ref{eq.exonsager_J1_he}), we can write
the extended Onsager relation Eq.~(\ref{eq.exonsager_J2_he}) and $\dot{Q}_c=J_2-P=J_3$ by using $J_1$ as
\begin{eqnarray}
&&J_2=\frac{L_{21}}{L_{11}}J_1+L_{22}(1-q^2)X_2-\gamma_h {J_1}^2,\label{eq.J2_by_J1_he}\\
&&J_3=\frac{L_{21}T_c}{L_{11}T_h}J_1+L_{22}(1-q^2)X_2+\gamma_c {J_1}^2,
 \label{eq.J3_by_J1_he}
\end{eqnarray}
respectively, where we define a constant meaning the strength of the dissipation into the cold heat reservoir $\gamma_c$ as
\begin{eqnarray}
\gamma_c \equiv \frac{T_c}{L_{11}}-\gamma_h >0,
\end{eqnarray}
assuming its positivity.
The meaning of each term in Eqs.~(\ref{eq.J2_by_J1_he}) and (\ref{eq.J3_by_J1_he}) can be highlighted by 
expressing the entropy production rate $\dot{\sigma}=-\frac{J_2}{T_h}+\frac{J_3}{T_c}$ by using them~\cite{yuki12a}:
\begin{eqnarray}
\dot{\sigma}=L_{22}(1-q^2)X_2^2+\frac{\gamma_h}{T_h}J_1^2+\frac{\gamma_c}{T_c}J_1^2,\label{eq.entropy_production_he}
\end{eqnarray}
where it is always assured to be non-negative from Eqs.~(\ref{eq.restrict_onsager_coeffi_he}) and (\ref{eq.q_he}), and the non-negativity of $\gamma_h>0$ and $\gamma_c>0$.
From Eq.~(\ref{eq.entropy_production_he}), we find that the first terms in Eqs.~(\ref{eq.J2_by_J1_he}) and (\ref{eq.J3_by_J1_he}) 
mean the reversible heat transports that do not contribute to the entropy production rate.
The second terms mean the steady heat leaks from the hot heat reservoir to the cold heat reservoir, which vanish under $|q|=1$ called the tight coupling. 
This tight-coupling condition in the MNLIT model assures that the heat fluxes $J_2$ and $J_3$ vanish simultaneously in the quasistatic limit $J_1\to 0$, playing a similar role in the LIT model.
The third terms mean the dissipations into the heat reservoirs due to the finite-time operation of the heat engine.
The power output $P=F\dot{x}=-J_1X_1T_c$ is also expressed by using $J_1$ as
\begin{eqnarray}
P=\frac{L_{12}}{L_{11}}\eta_{\rm C}J_1-\frac{T_c}{L_{11}}J_1^2.\label{eq.P_he}
\end{eqnarray}
Instead of the extended Onsager relations Eqs.~(\ref{eq.exonsager_J1_he}) and (\ref{eq.exonsager_J2_he}),
we can describe the heat engine by using Eqs.~(\ref{eq.J2_by_J1_he}) and (\ref{eq.J3_by_J1_he}).

Under given Onsager coefficients $L_{ij}$'s and the thermodynamic force $X_2$,
the working regime of the heat engines depends on $J_1$.
For the requirement of the positive power $P>0$, $J_1$ should be located in the following range:
\begin{eqnarray}
\begin{cases}
0<J_1< L_{12}X_2 & (0<L_{12}), \\
L_{12}X_2 <J_1<0 & (L_{12}<0). \label{eq.J1_range_he}
\end{cases}
\end{eqnarray}

The efficiency $\eta$ of the heat engine in our minimally nonlinear irreversible model is expressed as
\begin{eqnarray}
\eta=\frac{P}{J_2}=\frac{\frac{L_{12}}{L_{11}}\eta_{\rm C}J_1-\frac{T_c}{L_{11}}J_1^2}{\frac{L_{21}}{L_{11}}J_1+L_{22}(1-q^2)X_2-\gamma_h {J_1}^2},\label{eq.def_eta}
\end{eqnarray}
by using Eqs.~(\ref{eq.J2_by_J1_he}) and (\ref{eq.P_he}).

\subsection{Minimally nonlinear irreversible model for RE}\label{s2b}
As well as in the case of HE, 
we start from the entropy production rate $\dot{\sigma}$ of the total system (the refrigerator and the heat reservoirs). 
Because the internal state of the refrigerator comes back to the original state after one cycle for cyclic refrigerators 
or it remains unchanged for steady-state refrigerators, 
$\dot{\sigma}$ can be written by the sum of the entropy-change rate of the heat reservoirs as
$\dot{\sigma}=\dot{S}_h+\dot{S}_c$ according to the local equilibrium hypothesis~\cite{callen}, where $S_i$ ($i=h, c$) denotes the equilibrium entropy of the heat reservoir.
It can be written as
\begin{eqnarray}
\dot{\sigma}=\dot{S}_h+\dot{S}_c=\dot{U}_h\frac{\partial S_h}{\partial U_h}+\dot{U}_c\frac{\partial S_c}{\partial U_c}=\frac{\dot{Q}_h}{T_h}-\frac{\dot{Q}_c}{T_c},\label{eq.sigma_dot_re_1}
\end{eqnarray}
where we used $\frac{\partial S_i}{\partial U_i}=\frac{1}{T_i}$ with $U_i$ being the equilibrium internal-energy of the heat reservoir, and 
for RE, we denote by $\dot{Q}_c\equiv -\dot{U}_c$ the heat flux from the cold heat reservoir and by $\dot{Q}_h\equiv \dot{U}_h$ the heat flux into the hot heat reservoir, respectively, as is opposite to HE.
We also denote by $\dot{W}\equiv P$ the power injection. 
Then, from $\dot{Q}_h=P+\dot{Q}_c=F\dot{x}+\dot{Q}_c$
with $F$ and $x$ a generalized external force and its conjugate variable, respectively,
Eq.~(\ref{eq.sigma_dot_re_1}) can be rewritten as 
\begin{eqnarray}
\dot{\sigma}=\frac{F\dot{x}}{T_h}+\dot{Q}_c\left(\frac{1}{T_h}-\frac{1}{T_c}\right)\equiv J_1X_1+J_2X_2.
\end{eqnarray}
It naturally leads us to define the thermodynamic flux $J_1\equiv \dot{x}$ 
(the motion speed of the refrigerator)
conjugate to the thermodynamic force $X_1 \equiv F/T_h$, and the other thermodynamic flux $J_2\equiv \dot{Q}_c$ (the heat flux from the cold heat reservoir) conjugate to the other thermodynamic force $X_2\equiv 1/T_h-1/T_c$,
where these quantities are expressed by using the thermodynamic extensive and intensive parameters of the equilibrium heat reservoirs~\cite{callen}.

To establish our election of the thermodynamic fluxes and forces for refrigerators, 
we write the entropy production rate $\dot{\sigma}$ of the refrigerator as a function of $\dot Q_c$ and $\dot W$, thus incorporating in the formalism in a natural way the specific job of the refrigerator system (the extracted cooling power of the low-temperature reservoir $\dot Q_c$ as a consequence of the input of an external power $\dot W$). 
While it is more intuitive from a thermodynamic point of view that $\dot{\sigma}$ is written in terms of the specific job of each thermodynamic device in this way, an alternative starting point may be to express $\dot{\sigma}$ of the refrigerator in terms of $\dot W$ and $\dot Q_h$, as for heat engines. 
If this is done the thermodynamic fluxes and forces are exactly the same as those obtained for a heat engine, but it does not change the main results.

Then, as is similar to the heat engines in Sec.~II A, our minimally nonlinear irreversible refrigerator assumes the following extended Onsager relations between the thermodynamic fluxes $J_i$'s and forces $X_i$'s~\cite{yuki13a} [see Fig.~\ref{heat_device} (b)]:
\begin{eqnarray}
&&J_1=\dot{x}=L_{11}X_1+L_{12}X_2,\label{eq.exonsager_J1_re}\\
&&J_2=\dot{Q}_c=L_{21}X_1+L_{22}X_2-\gamma_c J_1^2,\label{eq.exonsager_J2_re}
\end{eqnarray}
where $L_{ij}$'s are the Onsager coefficients satisfying the reciprocal relation $L_{12}=L_{21}$.
The nonlinear term in $J_2$ expresses the dissipation into the cold heat reservoir caused by the finite-time motion of the refrigerator
and $\gamma_c$ ($>0$) is a constant meaning the strength of the dissipation.
In the absence of the nonlinear term, Eqs.~(\ref{eq.exonsager_J1_re}) and (\ref{eq.exonsager_J2_re}) recover the usual linear Onsager relations in LIT~\cite{vbroeck05}.
In LIT, the entropy production rate $\dot{\sigma}=J_1X_1+J_2X_2$ becomes the quadratic form in $X_i$'s and its non-negativity leads to the following restriction on the Onsager coefficients $L_{ij}$'s as
\begin{eqnarray}
L_{11}\ge 0, \ L_{22}\ge 0, \ L_{11}L_{22}-L_{12}^2\ge0.\label{eq.restrict_onsager_coeffi_re}
\end{eqnarray}
Although our extended Onsager relation Eq.~(\ref{eq.exonsager_J2_re}) includes the additional nonlinear term,
we assume that the restriction Eq.~(\ref{eq.restrict_onsager_coeffi_re}) still holds for our $L_{ij}$'s.

By changing the variable from $X_1$ to $J_1$ by using Eq.~(\ref{eq.exonsager_J1_re}), we can write the extended Onsager relation Eq.~(\ref{eq.exonsager_J2_re}) and $\dot{Q}_h=J_2+P\equiv J_3$ by using $J_1$ as
\begin{eqnarray}
&&J_2=\frac{L_{21}}{L_{11}}J_1+L_{22}(1-q^2)X_2-\gamma_c
 {J_1}^2,\label{eq.J2_by_J1_re}\\
&&J_3=\frac{L_{21}T_h}{L_{11}T_c}J_1+L_{22}(1-q^2)X_2+\gamma_h {J_1}^2,
 \label{eq.J3_by_J1_re}
\end{eqnarray}
respectively, where we define a constant meaning the strength of the dissipation into the hot heat reservoir $\gamma_h$ as
\begin{eqnarray}
\gamma_h \equiv \frac{T_h}{L_{11}}-\gamma_c >0,\label{eq.restrict_gh_ref}
\end{eqnarray}
assuming its positivity and the coupling strength parameter $q$ as
\begin{eqnarray}
q\equiv \frac{L_{12}}{\sqrt{L_{11}L_{22}}}\ \ (|q|\le 1).\label{eq.q_re}
\end{eqnarray}
Meaning of each term in Eqs.~(\ref{eq.J2_by_J1_re}) and (\ref{eq.J3_by_J1_re}) can be highlighted by 
expressing the entropy production rate $\dot{\sigma}=\frac{J_3}{T_h}-\frac{J_2}{T_c}$ by using them~\cite{yuki13a}:
\begin{eqnarray}
\dot{\sigma}=L_{22}(1-q^2)X_2^2+\frac{\gamma_h}{T_h}J_1^2+\frac{\gamma_c}{T_c}J_1^2,\label{eq.entropy_production_re}
\end{eqnarray}
where it is always assured to be non-negative from Eqs.~(\ref{eq.restrict_onsager_coeffi_re}) and (\ref{eq.q_re}), and the non-negativity of $\gamma_h>0$ and $\gamma_c>0$.
The expression of the entropy production rate Eq.~(\ref{eq.entropy_production_re}) for the refrigerator agrees with Eq.~(\ref{eq.entropy_production_he}) 
for the heat engines, presenting a unified description of heat devices in our MNLIT models.
From Eq.~(\ref{eq.entropy_production_re}), we find that the first terms in Eqs.~(\ref{eq.J2_by_J1_re}) and (\ref{eq.J3_by_J1_re}) mean the reversible heat transports that do not contribute to the entropy production rate.
The second terms mean the steady heat-leaks from the hot heat reservoir to the cold heat reservoir, which vanish under the tight-coupling condition $|q|=1$. 
Under this tight-coupling condition in the MNLIT model, the heat fluxes $J_2$ and $J_3$ vanish simultaneously in the quasistatic limit $J_1\to 0$, 
as it happens in the LIT model.
The third terms mean the dissipations into the heat reservoirs due to the finite-time operation of the refrigerator.
The power injection $P=F\dot{x}=J_1X_1T_h$ is also expressed by using $J_1$ as
\begin{eqnarray}
P=\frac{L_{12}}{L_{11}\varepsilon_{\mathrm{C}}}J_1+\frac{T_h}{L_{11}}J_1^2.\label{eq.P_re}
\end{eqnarray}
Instead of the extended Onsager relations Eqs.~(\ref{eq.exonsager_J1_re}) and (\ref{eq.exonsager_J2_re}),
we can describe the refrigerator by using Eqs.~(\ref{eq.J2_by_J1_re}) and (\ref{eq.J3_by_J1_re}).

Under given Onsager coefficients $L_{ij}$'s and the thermodynamic force $X_2$, the working regime of the refrigerators depends on $J_1$.
For the requirement of the positive cooling power $J_2>0$, $J_1$ should be located in the following range:
\begin{eqnarray}
\frac{L_{21}-\sqrt{D}}{2L_{11}\gamma_c}<J_1< \frac{L_{21}+\sqrt{D}}{2L_{11}\gamma_c}, \label{eq.J1_range_re}
\end{eqnarray}
where the discriminant $D$, which should be positive, is given by
\begin{eqnarray}
D\equiv L_{21}^2+4L_{11}^2L_{22}\gamma_c (1-q^2)X_2>0.\label{eq.discriminant}
\end{eqnarray}
Under the tight-coupling condition $|q|=1$, Eq.~(\ref{eq.J1_range_re}) reduces to
\begin{eqnarray}
\begin{cases}
0< J_1 <\frac{L_{21}}{\gamma_c L_{11}} & (L_{12}>0),\\
\frac{L_{21}}{\gamma_c L_{11}} < J_1 <0 & (L_{12}<0),
\end{cases}
\end{eqnarray}
showing that the quasistatic limit $J_1\to 0$ is included.
In contrast, under the nontight-coupling condition $|q|<1$, such quasistatic limit is not included and $J_1$ must be a finite value as in Eq.~(\ref{eq.J1_range_re}) for $J_2$ to be positive.
Intuitively, this constraint is necessary for the cooling effect to overcome the steady heat-leak effect. 
In addition, from Eq.~(\ref{eq.discriminant}), we also have a constraint on $\gamma_c$ under the non-tight-coupling condition $|q|<1$ as
\begin{eqnarray}
\gamma_c < -\frac{q^2}{4(1-q^2)L_{11}X_2}\equiv \gamma_c^+.\label{eq.gcplus}
\end{eqnarray}
This constraint is also related to the positivity of the cooling power under $|q|<1$:
even when the refrigerator operates at a finite rate,
large enough dissipation into the cold heat reservoir can also violate the positive cooling power.
Combining Eq.~(\ref{eq.gcplus}) with Eq.~(\ref{eq.restrict_gh_ref}), we obtain the following constraint on $\gamma_c$ depending on the parameter values as
\begin{eqnarray}
\begin{cases}
0< \gamma_c <\frac{T_h}{L_{11}} & \left(\gamma_c^+ \ge \frac{T_h}{L_{11}}, {\rm i.e.,} \ \tau \ge \frac{1}{\frac{q^2}{4(1-q^2)}+1}\right), \\
0<\gamma_c<\gamma_c^+ & \left(\gamma_c^+ < \frac{T_h}{L_{11}}, {\rm i.e.,} \ \tau < \frac{1}{\frac{q^2}{4(1-q^2)}+1}\right).\label{eq.gc_range}
\end{cases}
\end{eqnarray}
As is clear from Eq.~(\ref{eq.J1_range_he}), such restriction does not exist in the heat engines. However it plays a key role in the behavior of the optimum performance regimes of RE (see Sec. IV. B).

The COP $\varepsilon$ of the refrigerator in our minimally nonlinear irreversible model is expressed as
\begin{eqnarray}
\varepsilon=\frac{J_2}{P}=\frac{\frac{L_{21}}{L_{11}}J_1+L_{22}(1-q^2)X_2-\gamma_c
 {J_1}^2}{\frac{L_{12}}{L_{11}\varepsilon_{\mathrm{C}}}J_1+\frac{T_h}{L_{11}}J_1^2},\label{eq.def_cop}
\end{eqnarray}
by using Eqs.~(\ref{eq.J2_by_J1_re}) and (\ref{eq.P_re}).

\section{Global performance characteristics}
We consider global performance characteristics of the minimally nonlinear irreversible heat devices
described by Eqs.~(\ref{eq.J2_by_J1_he}) and (\ref{eq.J3_by_J1_he}) 
for HE or Eqs.~(\ref{eq.J2_by_J1_re}) and (\ref{eq.J3_by_J1_re}) for RE.

\subsection{HE: Performance characteristics}
\begin{figure*}
\includegraphics[scale=0.85]{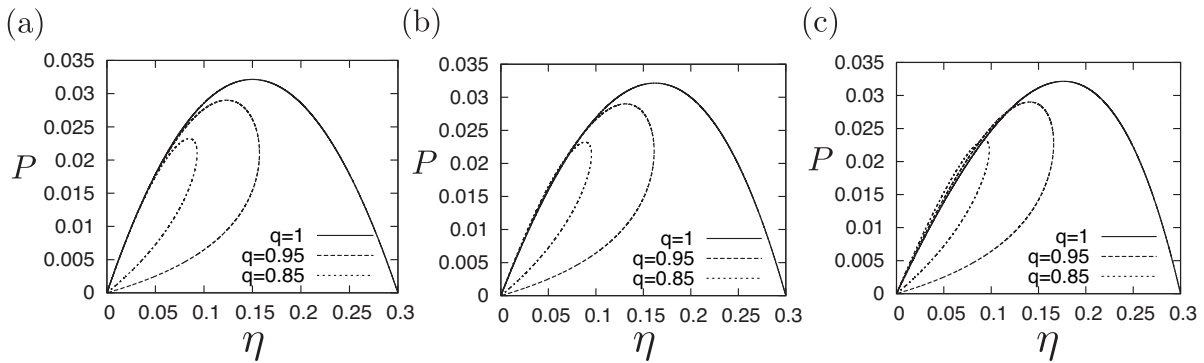}
\caption{Power-efficiency ($P$--$\eta$) curve for the minimally nonlinear irreversible heat engine under various coupling strengths (solid line for $q=1$, dashed line for $q=0.95$, and dotted line for $q=0.85$): 
(a) asymmetric dissipation ($\gamma_h=0.001$ and $\gamma_c=0.699$), (b) symmetric dissipation ($\gamma_h=\gamma_c=0.35$), and (c) asymmetric dissipation ($\gamma_h=0.699$ and $\gamma_c=0.001$). 
We used $L_{11}=L_{22}=1$, $T_h=1$, and $T_c=0.7$.
The Carnot efficiency is $\eta_{\rm C}=0.3$.}\label{gpc_p_eta_g}
\end{figure*}
\begin{figure*}
\includegraphics[scale=0.85]{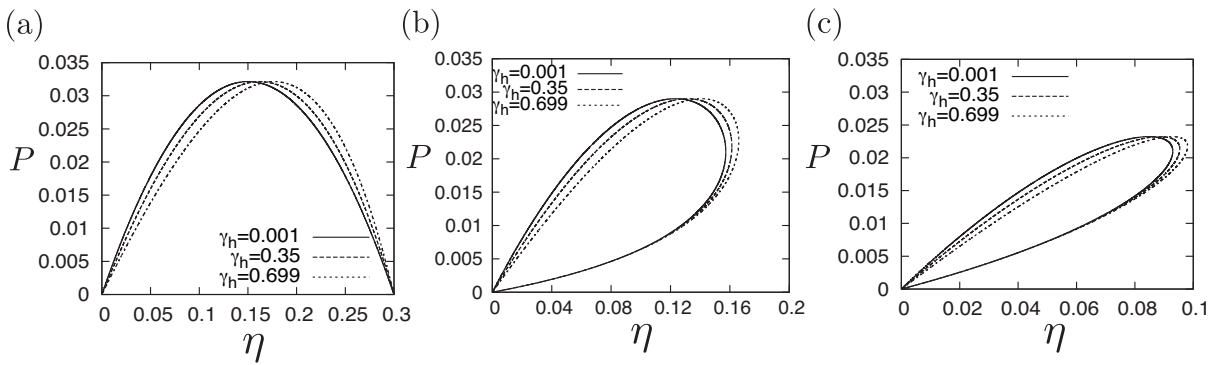}
\caption{Power-efficiency ($P$--$\eta$) curve for the minimally nonlinear irreversible heat engine under various dissipation regimes (solid line for $\gamma_h=0.001$, dashed line for $\gamma_h=0.35$, and dotted line for $\gamma_h=0.699$): 
(a) $q=1$, (b) $q=0.95$, and (c) $q=0.85$.
We used $L_{11}=L_{22}=1$, $T_h=1$, and $T_c=0.7$.
The Carnot efficiency is $\eta_{\rm C}=0.3$.}\label{gpc_p_eta_q}
\end{figure*}
In our model, $|q|=1$ means a perfect, tight-coupling condition for the internal degrees of freedom, so that the heat-leak terms proportional to the thermodynamic force $X_2$ in Eqs.~(\ref{eq.J2_by_J1_he}) and (\ref{eq.J3_by_J1_he}) do not play any role. Thus, the 
thermodynamic fluxes $J_{2}$ and $J_{3}$ depend only on the thermodynamic flux $J_{1\,}$ and the dissipation constants $\gamma_i$'s. 
The resulting power-efficiency plots for HE [Figs.~\ref{gpc_p_eta_g}(a)--\ref{gpc_p_eta_g}(c) for $|q|=1$] are parabolic shaped defined between the two null-power states corresponding to a stalled state~\cite{vbroeck05,seifert08} 
where the power vanishes due to too quick operation of the heat engine and the Carnot efficiency state realized in the quasistatic limit $J_1\to 0$, independently of the dissipation constants.
These figures exhibit the characteristics common to general Carnot-like models when the heat leak is absent, and the irreversibilities are thus limited to external coupling between the working system and external heat reservoirs with adequate heat transfer laws (endoreversible limit~\cite{lchen99,wu04,jchen01,luisarias13,gordon00}). Only if additionally the heat transfer law is linear, the CA efficiency emerges in FTT~\cite{curzon75}. 
However, this is not the case in our model, where this particular value is realized only as a limiting case. Later we will come back to this particular point in Sec.~IV.

For $|q|<1$, the thermodynamic fluxes $J_2$ and $J_3$ 
also depend on an additional direct heat transfer between the hot and cold heat reservoirs, which is proportional to $X_2$ as in Eqs.~(\ref{eq.J2_by_J1_he}) and (\ref{eq.J3_by_J1_he}). Now, the resulting power-efficiency ($P$--$\eta$) plots are loop shaped (as those obtained in the irreversible Carnot-like FTT models~\cite{jchen01}) with near but non-coincident maximum power and maximum efficiency points. As the degree of the heat leak increases (i.e., as $|q|$ decreases) [Figs.~\ref{gpc_p_eta_g}(a)--\ref{gpc_p_eta_g}(c)], the distance between maximum power and maximum efficiency points becomes smaller determining more closed loops. This loop-shaped behavior is explained as follows: when $|q|< 1$,
the heat-leak effect steadily remains even in the quasistatic limit $J_1 \to 0$, which implies $\eta \to 0$ from Eq.~(\ref{eq.def_eta}).
In contrast, even when the magnitude of $J_1$ becomes too large,
the efficiency (as well as the power) becomes $0$ at the stalled state $J_1=L_{12}X_2$~\cite{vbroeck05,seifert08}.
Therefore the optimum $J_1$ that maximizes $\eta$ must exist somewhere between these extreme points.

The explicit influence of the dissipation constants $\gamma_i$'s on the global performance characteristics is better visualized from Figs.~\ref{gpc_p_eta_q}(a)--\ref{gpc_p_eta_q}(c). It is clearly observed that at any fixed power (including the maximum power point) and any $q$ value, the efficiency increases as $\gamma_{h}$ increases. This result generalizes the result reported by Apertet {\it et al.}~\cite{apertet12a,apertet12b} for a tightly coupled thermoelectric generator to any coupling case. As these authors of Refs.~\cite{apertet12a,apertet12b} argue, the heat released at the hot heat reservoir can eventually be recycled by the hot isothermal step, and thus a preferential dissipation into this side provokes increase of the efficiency at any fixed power. 
Actually, this mechanism works for any $q$ value: we can obtain the $\eta=\eta(P)$ curve explicitly by combining Eqs.~(\ref{eq.P_he}) and (\ref{eq.def_eta}) as
\begin{eqnarray}
\eta(P)=\frac{P}{\frac{q^2L_{22}X_2C_h}{2}+L_{22}(1-q^2)X_2-\gamma_h \frac{L_{12}^2X_2^2C_h^2}{4}},\label{eq.eta_p_curve}
\end{eqnarray}
where $C_h$ is defined as
\begin{eqnarray}
C_h \equiv 1\pm \sqrt{1-\frac{4PT_c}{q^2L_{22}\eta_{\rm C}^2}},
\end{eqnarray}
and the sign $+$ ($-$) corresponds to a branch for the working regimes from the stalled-state point (quasistatic limit) to the maximum power point. 
From the curve Eq.~(\ref{eq.eta_p_curve}), it is obvious that the efficiency increases as $\gamma_h$ increases at any fixed power and for any $q$ value.

\subsection{RE: Performance characteristics}

\begin{figure*}
\includegraphics[scale=0.72]{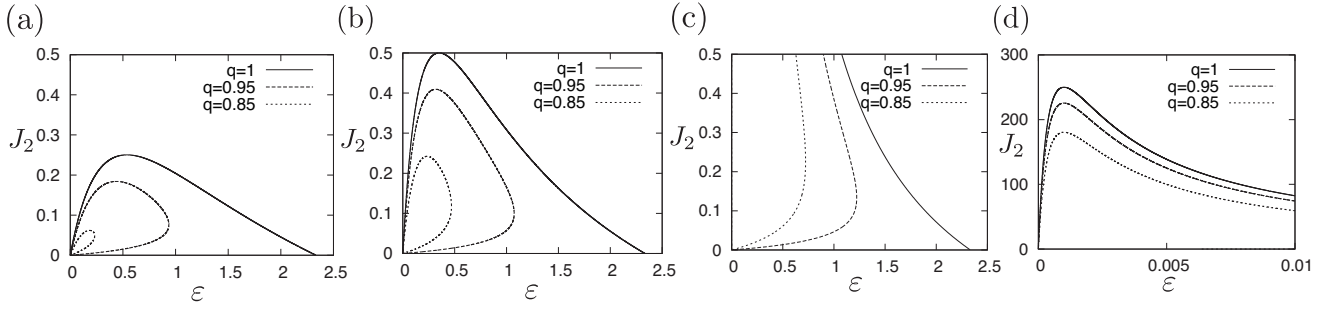}
\caption{Cooling power-COP ($J_2$--$\varepsilon$) curve for the minimally nonlinear irreversible refrigerator under various
coupling strengths (solid line for $q=1$, dashed line for $q=0.95$, and dotted line for $q=0.85$): (a) asymmetric dissipation ($\gamma_h=0.001$ and $\gamma_c=0.999$), (b) symmetric dissipation ($\gamma_h=\gamma_c=0.5$), and (c) asymmetric dissipation ($\gamma_h=0.999$ and $\gamma_c=0.001$). (d) shows a closer inspection of (c) in the small $\varepsilon$-range for better understanding of the shape of $J_2$--$\varepsilon$ curve (open curve for $q=1$ and closed curve for $q=0.95, 0.85$). Diverging behavior of $J_2$ in the limit of $\gamma_c \to 0$ is confirmed. 
We used $L_{11}=L_{22}=1$, $T_h=1$, and $T_c=0.7$.
The Carnot COP is $\varepsilon_{\rm C}\simeq 2.33$.}
\label{gpc_j2_cop_g}
\end{figure*}
\begin{figure*}
\includegraphics[scale=0.85]{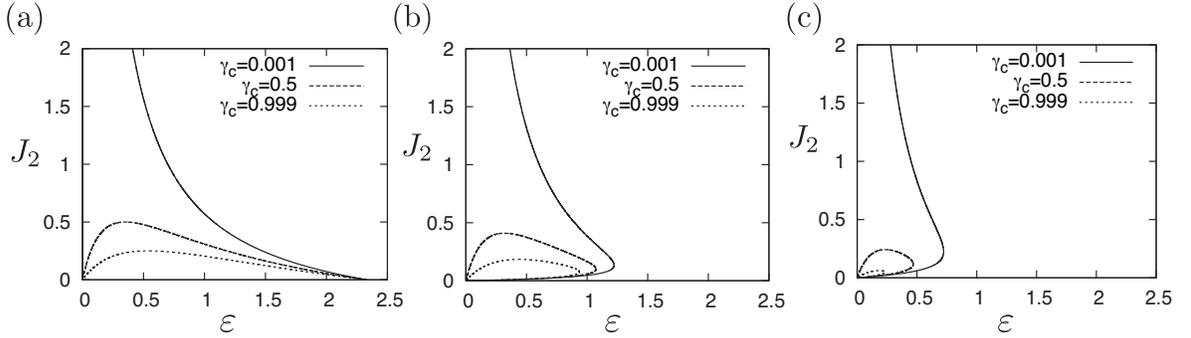}
\caption{Cooling power-COP ($J_2$--$\varepsilon$) curve for the minimally nonlinear irreversible refrigerator under various dissipation regimes (solid line for $\gamma_c=0.001$, dashed line for $\gamma_c=0.5$, 
and dotted line for $\gamma_c=0.999$): 
(a) $q=1$, (b) $q=0.95$, and (c) $q=0.85$.
We used $L_{11}=L_{22}=1$, $T_h=1$, and $T_c=0.7$.
The Carnot COP is $\varepsilon_{\rm C}\simeq 2.33$.}\label{gpc_j2_cop_q}
\end{figure*}

Following the same methodology as for HE above, we will analyze the 
cooling power-COP ($J_2$--$\varepsilon$) plots. See Figs.~\ref{gpc_j2_cop_g}(a)--\ref{gpc_j2_cop_g}(d).

Again, we clearly observe open curves under the tight-coupling condition $|q|=1$, which are similar to the characteristics of the endoreversible models~\cite{lchen99,wu04,jchen01,luisarias13,gordon00}, where the maximum cooling power is realized at a finite rate while the maximum COP is realized at the zero cooling power (quasistatic limit).
In contrast, under the non-tight-coupling condition $|q|<1$, the maximum COP is no longer realized at the zero cooling power but at a finite cooling power because in this case the COP at the zero cooling power (quasistatic limit) becomes zero due to a steady heat leak. 
Then we observe loop-shaped curves with near but non-coincident optimum values for both of the cooling power and COP.
As $|q|$ progressively decreases, this tendency becomes prominent and 
both of the cooling power and COP become smaller and their behaviors are also strongly 
modulated by the values of the dissipation constants $\gamma_i$'s.
This loop-shaped behavior is explained as in the HE case above: when $|q|< 1$, $J_1$ takes values in a bounded interval as in Eq.~(\ref{eq.J1_range_re}) for the cooling power $J_2$ to be positive.
At both ends in Eq.~(\ref{eq.J1_range_re}), $\varepsilon$ becomes $0$ because $J_2$ vanishes there [see Eq.~(\ref{eq.def_cop})].
Therefore the optimum $J_1$ that maximizes $\varepsilon$ must exist somewhere between these extreme points. We also see that, in the limit of $\gamma_c \to 0$, the maximum cooling power shows diverging behavior for all $q$'s, 
which has previously been pointed out by Apertet {\it et al.} for the tight-coupling case $|q|=1$ in the thermoelectric generator~\cite{apertet13a}.
Similar behavior has also been reported for the cooling power at the maximum $\chi$ condition for minimally nonlinear irreversible refrigerators~\cite{yuki13a}.

The explicit influence of the dissipation constants $\gamma_i$'s on the global performance characteristics can be analyzed more clearly from Figs.~\ref{gpc_j2_cop_q}(a)--\ref{gpc_j2_cop_q}(c). 
For $|q|=1$ in Fig.~\ref{gpc_j2_cop_q}(a),  a preferential dissipation into the cold heat reservoir ($\gamma_c\rightarrow \frac{T_h}{L_{11}}$) provokes a decreasing of the cooling power at any fixed COP, but the COP at maximum cooling power monotonically increases.
This counterintuitive behavior was also reported in Ref.~\cite{apertet13a} for the tight-coupling case $|q|=1$ in the thermoelectric generator. 
For more realistic situations with $|q|<1$ as in Figs.~\ref{gpc_j2_cop_q}(b) and \ref{gpc_j2_cop_q}(c), 
we can see that the preferential dissipation into the cold heat reservoir generally induces the smaller loop.
Interestingly, we can also see that the COP at maximum cooling power can show a non-monotonic behavior with respect to the strength of the dissipation as in Fig.~\ref{gpc_j2_cop_q}(c) ($q=0.85$) in contrast to the monotonic behavior as in \ref{gpc_j2_cop_q}(a) ($|q|=1$) and \ref{gpc_j2_cop_q}(b) ($q=0.95$).
This behavior of the COP at maximum cooling power will be discussed in detail in Sec.~IV B.
\section{Optimized regimes}
We analyze in detail the optimized performance regimes of the maximum efficiency and the efficiency at maximum power for HE and those of the maximum COP and the COP at maximum cooling power for RE found in the analysis of the global performance regimes in Sec.~III.
\subsection{Optimized regimes: HE}
\begin{figure*}
\includegraphics[scale=1.0]{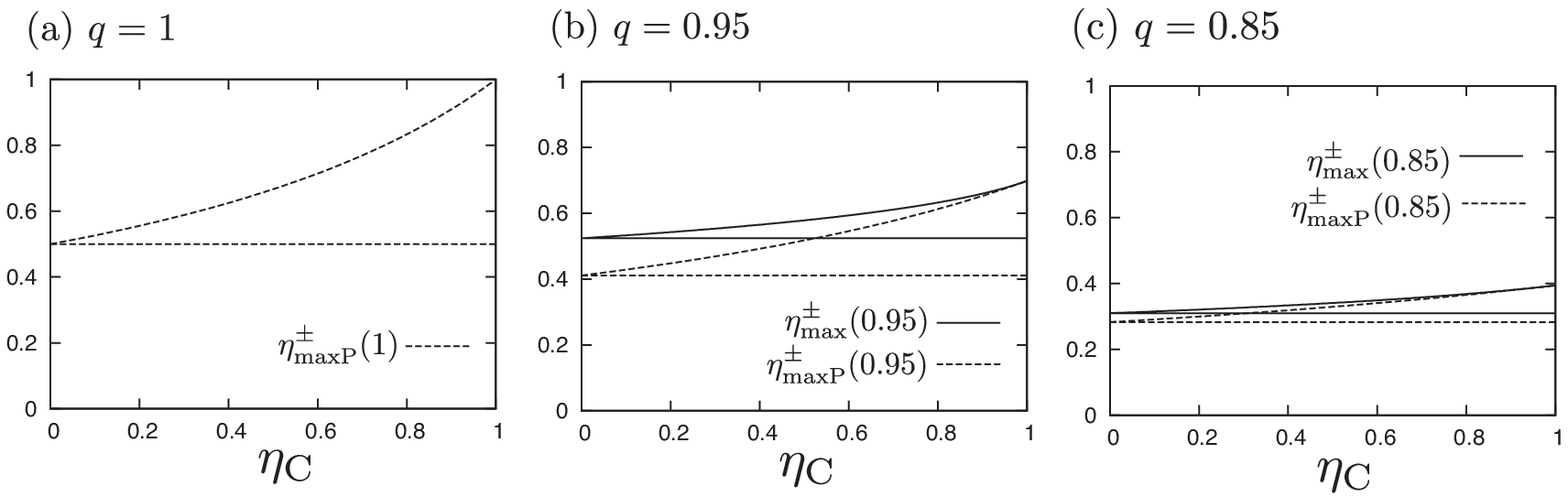}
\caption{Normalized upper bounds $\eta_{\rm max}^+(q)/\eta_{\rm C}$ in Eq.~(\ref{eq.etamaxup}) and $\eta_{\rm maxP}^+(q)/\eta_{\rm C}$ in Eq.~(\ref{eq.eta_maxpow_upper}), and lower bounds $\eta_{\rm max}^-(q)/\eta_{\rm C}$ in Eq.~(\ref{eq.etamaxlow}) and $\eta_{\rm maxP}^-(q)/\eta_{\rm C}$ in Eq.~(\ref{eq.eta_maxpow_lower}) as a function of $\eta_{\rm C}$: (a) $q=1$, (b) $q=0.95$, and (c) $q=0.85$. $\eta_{\rm max}^+(1)=\eta_{\rm max}^-(1)=\eta_{\rm C}$ 
[see Eqs.~(\ref{eq.etamaxlow}) and (\ref{eq.etamaxup})].
}\label{lower_and_upper_bound_eta_max}
\end{figure*}

By using the definition of $\eta$ in Eq.~(\ref{eq.def_eta}) and solving $\partial \eta/\partial J_1=0$, we obtain
the maximum efficiency $\eta_{\rm max}(q, \gamma_h/\gamma_c)$ explicitly as:
\begin{widetext}
\begin{eqnarray}
\eta_{\rm max}\left(q, \frac{\gamma_h}{\gamma_c} \right)=\frac{\eta_{\rm C}+\frac{\left(1+\frac{\gamma_c}{\gamma_h}\right)(1-q^2)\eta_{\rm C}}{q^2\left(1-\eta_{\rm C}+\frac{\gamma_c}{\gamma_h}\right)}\left(1-\sqrt{1+\frac{q^2\left(1-\eta_{\rm C}+\frac{\gamma_c}{\gamma_h}\right)}{(1-q^2)\left(1+\frac{\gamma_c}{\gamma_h}\right)}}\right)}{1-\frac{1-\eta_{\rm C}+\frac{\gamma_c}{\gamma_h}}{1+\frac{\gamma_c}{\gamma_h}}\frac{1}{\left(1-\sqrt{1+\frac{q^2\left(1-\eta_{\rm C}+\frac{\gamma_c}{\gamma_h}\right)}{(1-q^2)\left(1+\frac{\gamma_c}{\gamma_h}\right)}}\right)}+\frac{(1-q^2)\eta_{\rm C}}{q^2\left(1+\frac{\gamma_c}{\gamma_h}\right)}\left(1-\sqrt{1+\frac{q^2\left(1-\eta_{\rm C}+\frac{\gamma_c}{\gamma_h}\right)}{(1-q^2)\left(1+\frac{\gamma_c}{\gamma_h}\right)}}\right)}.\label{eq.eta_max_nontight}
\end{eqnarray}
\end{widetext}
This is a monotonically increasing function of $|q|$ and $\gamma_h/\gamma_c$~\cite{monotonicity}.
In the limit of the small temperature difference $\Delta T\to 0$, Eq.~(\ref{eq.eta_max_nontight}) can be expanded as
\begin{eqnarray}
\eta_{\rm max}\left(q, \frac{\gamma_h}{\gamma_c} \right)=\frac{\left(1-\sqrt{1-q^2}\right)^2}{q^2}\frac{\Delta T}{T}+O({\Delta T}^2),
\end{eqnarray}
where $T\equiv (T_h+T_c)/2$. Up to the first order of $\Delta T$, we have no $\gamma_h/\gamma_c$ dependence.
This expression has been previously obtained in the framework of LIT~\cite{cisneros08}.
In asymmetric dissipation limits $\gamma_h/\gamma_c \to 0$ and $\gamma_h/\gamma_c \to \infty$,
we find that Eq.~(\ref{eq.eta_max_nontight}) is bounded from the lower side by $\eta_{\rm max}^-(q)$ and the upper side by $\eta_{\rm max}^+(q)$,
which are given as
\begin{eqnarray}
\eta_{\rm max}^-(q) &&\equiv \eta_{\rm max}\left(q,\frac{\gamma_h}{\gamma_c} \rightarrow 0\right)\nonumber\\
&&=\frac{\left(1-\sqrt{1-q^2}\right)^2}{q^2}\eta_{\rm C},\label{eq.etamaxlow}
\end{eqnarray}
and
\begin{eqnarray}
&&\eta_{\rm max}^+(q)\equiv \eta_{\rm max}\left(q,\frac{\gamma_h}{\gamma_c} \rightarrow \infty \right)=\nonumber\\ 
&&\frac{\sqrt{\frac{1-q^2\eta_{\rm C}}{1-q^2}}\left(q^2\eta_{\rm C}^2+(q^2-2)\eta_{\rm C}\right)+2\eta_{\rm C}(1-q^2\eta_{\rm C})} {\sqrt{\frac{1-q^2\eta_{\rm C}}{1-q^2}}\left((3q^2-2)\eta_{\rm C}-q^2\right)+2\eta_{\rm C}(1-q^2\eta_{\rm C})},\label{eq.etamaxup}
\end{eqnarray}
respectively.
In the limit of $|q|\to 1$, $\eta^{\pm}_{\rm max}(q) \to \eta_{\rm C}$ as expected,
but as the coupling strength decreases, these values drastically decrease showing the behavior plotted in Figs.~\ref{lower_and_upper_bound_eta_max}(b) and \ref{lower_and_upper_bound_eta_max}(c).

In the case of the symmetric dissipation $\gamma_h=\gamma_c$, we obtain
\begin{eqnarray}
\eta_{\rm max}^{\rm sym}(q)&&\equiv \eta_{\rm max}\left(q,\frac{\gamma_h}{\gamma_c}=1\right)\\\nonumber
&&=\frac{\eta_{\rm C}+\frac{2(1-q^2)\eta_{\rm C}}{q^2(2-\eta_{\rm C})}H_1}{1-\frac{2-\eta_{\rm C}}{2}\frac{1}{H_1}+\frac{(1-q^2)\eta_{\rm C}}{2q^2}H_1},
\end{eqnarray}
where $H_1$ is defined as
\begin{eqnarray}
H_1\equiv 1-\sqrt{1+\frac{q^2(2-\eta_{\rm C})}{2(1-q^2)}}.
\end{eqnarray}

Among different optimization regimes, the efficiency at maximum power $\eta_{\rm maxP}$ has been playing an important role
for studies of traditional~\cite{esposito10a, apertet12a, apertet12b,tu12abc, jwang12a,bizarro,ito,jchen13a,broeck13a,norma10a}, stochastic~\cite{bekele04,tu08,seifert08,gaveau10,tu12a,seifert12,broeck12a,he13a}, and quantum~\cite{esposito09a,jwang12b,hyan12a,bekele12,rwang13a,jchen13b,allahv13} HE. 
The maximum power and the efficiency at the maximum power $\eta_{\rm maxP}$ of the present model, which were studied in Ref.~\cite{yuki12a} previously, are obtained by solving $\partial P/\partial J_1=0$:
\begin{eqnarray}
P_{\rm max}&&=\frac{q^2L_{22}\eta_{\rm C}^2}{4T_c},\\
\eta_{\rm maxP}\left(q, \frac{\gamma_h}{\gamma_c}\right)&&=
\frac{\eta_{\rm C}}{2} \frac{q^2}{2-q^2\left(1+\frac{\eta_{\rm C}}{2\left(1+\frac{\gamma_c}{\gamma_h}\right)}\right)},\label{remchig}
\end{eqnarray}
where we used the definition of $P$ in Eq.~(\ref{eq.P_he}) and $\eta$ in Eq.~(\ref{eq.def_eta}).
Equation.~(\ref{remchig}) is a monotonically increasing function of $|q|$ and $\gamma_h/\gamma_c$~\cite{yuki12a}.
The corresponding lower and upper bounds, and symmetric case are easily obtained in the limit of 
$\gamma_h/\gamma_c \to 0$, 
$\gamma_h/\gamma_c \to \infty$, and $\gamma_c/\gamma_h=1$, respectively:
\begin{eqnarray}
\eta^-_{\rm maxP}(q)\equiv \eta_{\rm maxP}\left(q,\frac{\gamma_h}{\gamma_c}\rightarrow 0\right)=\frac{\eta_{\rm C}}{2}\frac{q^2}{2-q^2},\label{eq.eta_maxpow_lower}
\end{eqnarray}
\begin{eqnarray}
\eta^{\rm sym}_{\rm maxP}(q)&&\equiv \eta_{\rm maxP}\left(q,\frac{\gamma_h}{\gamma_c}=1\right)\nonumber\\
&&=\frac{\eta_{\rm C}}{2}\frac{q^2}{2-q^2(1+\frac{\eta_{\rm C}}{4})},
\end{eqnarray}
\begin{eqnarray}
\eta^+_{\rm maxP}(q)&&\equiv \eta_{\rm maxP}\left(q,\frac{\gamma_h}{\gamma_c}\rightarrow \infty \right)\nonumber\\
&&=\frac{\eta_{\rm C}}{2}\frac{q^2}{2-q^2(1+\frac{\eta_{\rm C}}{2})}.\label{eq.eta_maxpow_upper}
\end{eqnarray}
The bounds $\eta^-_{\rm maxP}(q)$ and $\eta^+_{\rm maxP}(q)$ are also plotted in Figs.~\ref{lower_and_upper_bound_eta_max}(a)--\ref{lower_and_upper_bound_eta_max}(c) 
for the sake of comparison with the lower and upper bounds $\eta^-_{\rm max}(q)$ and $\eta^+_{\rm max}(q)$ of the maximum efficiency. 
Note that as the coupling strength progressively decreases due to the heat-leak increase [Fig.~\ref{lower_and_upper_bound_eta_max}(c)], 
the maximum efficiency and efficiency at the maximum power regimes tend to collapse in a unique inefficient performance regime as we have seen, for instance, in Fig.~\ref{gpc_p_eta_q}.

The following limits of Eq.~(\ref{remchig}) under the tight-coupling condition $|q|=1$ are especially interesting:
\begin{eqnarray}
\eta_{\rm maxP}^{-}=\frac{\eta_{\rm C}}{2}\,\,\left(\frac{\gamma_h}{\gamma_c}\to 0, \ |q|=1 \right),\label{remchilb}
\end{eqnarray}
\begin{eqnarray}
\eta_{\rm maxP}^{\rm sym}=\frac{2\eta_{\rm C}}{4-\eta_{\rm C}}\,\,\left(\frac{\gamma_h}{\gamma_c}=1, \ |q|=1 \right),\label{remchisyma}
\end{eqnarray}
\begin{eqnarray}
\eta_{\rm maxP}^{+}=\frac{\eta_{\rm C}}{2-\eta_{\rm C}}\,\,\left(\frac{\gamma_h}{\gamma_c}\to \infty, \ |q|=1 \right),\label{remchiub}
\end{eqnarray}
where Eqs.~(\ref{remchilb}) and (\ref{remchiub}) are previously obtained in the LD models~\cite{esposito10a}, and Eq.~(\ref{remchisyma}) may be comparable to the previous result obtained for a heat engine model 
under a left-right (spatially) symmetric condition in Ref.~\cite{esposito09a}.
We note that when $\eta_{\rm C}\rightarrow 0$ ($T_c \sim T_h$), $\eta_{\rm maxP}^{\rm sym}$ in Eq.~(\ref{remchisyma}) is expanded as
\begin{eqnarray}\label{remchisymb}
\eta_{\rm maxP}^{\rm sym}=\frac{2\eta_{\rm C}}{4-\eta_{\rm C}}\approx \frac{\eta_{\rm C}}{2}+\frac{\eta_{\rm C}^2}{8}+\frac{\eta_{\rm C}^3}{32}+\cdots,
\end{eqnarray}
which reproduces $\eta_{\rm CA}$~\cite{curzon75} up to the second order of $\eta_{\rm C}$:
\begin{equation}\label{remca}
\eta_{\rm CA}=1-\sqrt{1-\eta_{\rm C}}\approx \frac{\eta_{\rm C}}{2}+\frac{\eta_{\rm C}^2}{8}+\frac{\eta_{\rm C}^3}{16}+\cdots.
\end{equation}
Consequently, under the maximum power condition, our model optimized with respect to only one-parameter $J_1$ reproduces the lower and upper bounds of the low-dissipation HE model optimized with respect to two parameters in~\cite{esposito10a} and also reproduces the CA efficiency up to second order in the limit of the small temperature difference~\cite{esposito09a}.
\subsection {Optimized regimes: RE}
\begin{figure*}
\includegraphics[scale=1.0]{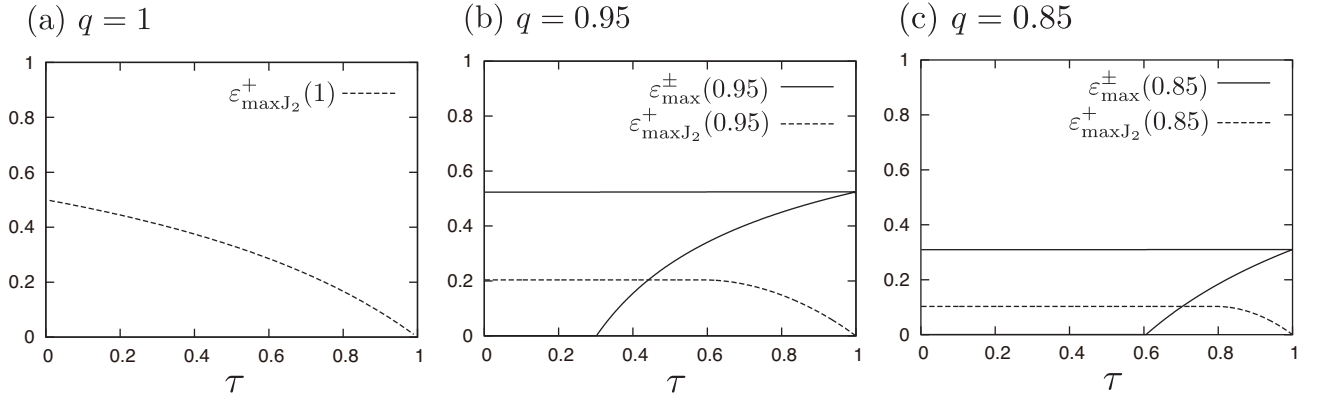}
\caption{Normalized upper bounds $\varepsilon_{\rm max}^+(q)/\varepsilon_{\rm C}$ in Eq.~(\ref{eq.maxcop_upper}) and $\varepsilon_{\rm maxJ_2}^+(q)/\varepsilon_{\rm C}$ in Eq.~(\ref{eq.copmaxj2_upper}), and lower bound $\varepsilon_{\rm max}^-(q)/\varepsilon_{\rm C}$ in Eq.~(\ref{eq.maxcop_lower}) as a function of $\tau$:
(a) $q=1$, (b) $q=0.95$, and (c) $q=0.85$.
$\varepsilon_{\rm max}^+(1)/\varepsilon_{\rm C}=\varepsilon_{\rm max}^-(1)/\varepsilon_{\rm C}=1$ in all $\tau$-range (see Eqs.~(\ref{eq.maxcop_upper}) and (\ref{eq.maxcop_lower})).
}\label{lower_and_upper_bound_cop_max}
\end{figure*}
By using the definition of $\varepsilon$ in Eq.~(\ref{eq.def_cop}) and solving $\partial \varepsilon/\partial J_1=0$, 
the maximum COP under the nontight-coupling condition $|q|<1$ is given as follows:
\begin{eqnarray}
\varepsilon_{\rm max}\left(q, \frac{\gamma_h}{\gamma_c}\right)=\frac{-q^2R_1 R_2+q^2R_1^2-\frac{(1-q^2)(1-\frac{1}{\tau})R_2^2}{1+\frac{\gamma_h}{\gamma_c}}}{q^2(1-\frac{1}{\tau})R_1R_2+(1-q^2)(1-\frac{1}{\tau})R_2^2},\label{eq.maxcop_re}
\end{eqnarray}
where $R_1$ and $R_2$ are defined as
\begin{eqnarray}
&&R_1 \equiv 1-\frac{1-\frac{1}{\tau}}{1+\frac{\gamma_h}{\gamma_c}},\\
&&R_2 \equiv 1+\sqrt{1+\frac{q^2R_1}{1-q^2}},
\end{eqnarray}
respectively.
Eq.~(\ref{eq.maxcop_re}) is a monotonically increasing function of $|q|$ and $\gamma_h/\gamma_c$~\cite{monotonicity}.
In the asymmetric dissipation limit $\gamma_h/\gamma_c \to \infty$, we obtain $R_1 \to 1$, $R_2 \to 1+\sqrt{1+q^2/(1-q^2)}$, and the upper bound:
\begin{eqnarray}
\varepsilon_{\rm max}^+(q)=\frac{q^2}{(1+\sqrt{1-q^2})^2}\varepsilon_{\rm C}.\label{eq.maxcop_upper}
\end{eqnarray}
We also obtain the lower bound as
\begin{widetext}
\begin{eqnarray}
\varepsilon_{\rm max}^-(q)=\label{eq.maxcop_lower}
\begin{cases}
\frac{\frac{q^2}{{\tau}^2}-\frac{q^2}{\tau}\left(1+\sqrt{1+\frac{q^2}{(1-q^2)\tau}}\right)-(1-q^2)(1-\frac{1}{\tau})\left(1+\sqrt{1+\frac{q^2}{(1-q^2)\tau}}\right)^2}{(1-\frac{1}{\tau})\frac{q^2}{\tau}\left(1+\sqrt{1+\frac{q^2}{(1-q^2)\tau}}\right)+(1-q^2)(1-\frac{1}{\tau})\left(1+\sqrt{1+\frac{q^2}{(1-q^2)\tau}}\right)^2} & \left(\frac{\gamma_h}{\gamma_c}\to 0 \ {\rm for} \ \tau \ge \frac{1}{\frac{q^2}{4(1-q^2)}+1}\right), \\
0 & \left(\frac{\gamma_h}{\gamma_c}\to \frac{\gamma_h}{\gamma_c^+} \ {\rm for} \ \tau < \frac{1}{\frac{q^2}{4(1-q^2)}+1}\right),
\end{cases}
\end{eqnarray}
\end{widetext}
depending on the parameter values corresponding to each case in Eq.~(\ref{eq.gc_range}).
$\varepsilon_{\rm max}^+(q)$ and $\varepsilon_{\rm max}^-(q)$ normalized by $\varepsilon_{\rm C}$ are plotted in Figs.~\ref{lower_and_upper_bound_cop_max}(b) and \ref{lower_and_upper_bound_cop_max}(c).

In the case of the symmetric dissipation $\gamma_h=\gamma_c$, we obtain 
\begin{eqnarray}
\varepsilon_{\rm max}^{\rm sym}(q)=\frac{q^2(1-\frac{1}{\tau})-q^2R_3-(1-q^2)R_3^2}{q^2(1-\frac{1}{\tau})R_3+2(1-q^2)R_3^2},
\end{eqnarray}
where 
\begin{eqnarray}
R_3 \equiv 1+\sqrt{1+\frac{q^2(1-\frac{1}{\tau})}{2(1-q^2)}}.
\end{eqnarray}

It is evident from Figs.~\ref{gpc_j2_cop_g} and \ref{gpc_j2_cop_q} that the COP at the maximum cooling power $\varepsilon_{\rm maxJ_2}$ is a well-defined optimum performance regime as is also suggested in~\cite{apertet13a}. 
Analytical derivation of the maximum cooling power $J_{2, {\rm max}}$ and the COP at the maximum cooling power $\varepsilon_{\rm maxJ_2}$ are easily done as $\partial J_2/\partial J_1=0$ by using Eq.~(\ref{eq.J2_by_J1_re}) and they read as
\begin{eqnarray}
J_{2, {\rm max}}&&=\frac{q^2L_{22}}{4\gamma_c L_{11}}+L_{22}(1-q^2)X_2,\label{eq.maxJ2} \\ 
\varepsilon_{\rm max J_2}\left(q, \frac{\gamma_h}{\gamma_c}\right)&&=\frac{q^2\varepsilon_{\rm C}-\frac{4(1-q^2)}{(1+\frac{\gamma_h}{\gamma_c})}}{2q^2+q^2\varepsilon_{\rm C}\left(1+\frac{\gamma_h}{\gamma_c}\right)}.\label{eq.cop_maxJ2}
\end{eqnarray} 
Eq.~(\ref{eq.cop_maxJ2}) is a monotonically increasing function of $|q|$, and a monotonically decreasing function of $\gamma_h/\gamma_c$ for $|q|=1$ while for $|q|\ne 1$ it depends as follows.  
By solving $\partial \varepsilon_{\rm maxJ_2}/\partial (\gamma_h/\gamma_c)=0$, we obtain the optimum dissipation ratio:  
\begin{eqnarray}
\left(\frac{\gamma_h}{\gamma_c}\right)_{\rm opt}=-1+\frac{4(1-q^2)}{q^2\varepsilon_{\rm C}} \left(1+\sqrt{1+\frac{q^2}{2(1-q^2)}}\right).\label{eq.opt_dissipation}
\end{eqnarray}
Because of the requirement $\left(\frac{\gamma_h}{\gamma_c}\right)_{\rm opt}>0$, we obtain a parameter range 
\begin{eqnarray}
\varepsilon_{\rm C} < \frac{4(1-q^2)\left(1+\sqrt{1+\frac{q^2}{2(1-q^2)}}\right)}{q^2} 
\end{eqnarray}
for this optimum value to exist. A non-monotonic behavior corresponding to this case can be seen in 
Fig.~\ref{gpc_j2_cop_q} (c) ($\varepsilon_{\rm maxJ_2}$ of $\gamma_c=0.5$ is the maximum in the 
three $\gamma_c$'s.), as mentioned in the last part of Sec. III B.
If this condition does not hold [e.g., the parameter values used in Fig.~\ref{gpc_j2_cop_q} (b)], Eq.~(\ref{eq.cop_maxJ2}) is a monotonically decreasing function of $\gamma_h/\gamma_c$~\cite{monotonicity} 
as is similar to the case of $|q|=1$.
Then the upper bound of $\varepsilon_{\rm maxJ_2}(q)$ depending on the parameter values is given as follows:
\begin{widetext}
\begin{eqnarray}
\varepsilon_{\rm maxJ_2}^+(q) &&=\label{eq.cop_maxJ2_optimum}
\begin{cases}
\frac{q^2\varepsilon_{\rm C}R_4}{2(1+R_4)\left(q^2+2(1-q^2)(1+R_4)\right)} \ \left(\frac{\gamma_h}{\gamma_c}=\left(\frac{\gamma_h}{\gamma_c}\right)_{\rm opt} \ {\rm for} \ \varepsilon_{\rm C} < \frac{4(1-q^2)\left(1+\sqrt{1+\frac{q^2}{2(1-q^2)}}\right)}{q^2}  \right),\\
\frac{q^2\varepsilon_{\rm C}-4(1-q^2)}{q^2(2+\varepsilon_{\rm C})}\ \left(\frac{\gamma_h}{\gamma_c}\to 0 \ {\rm for} \ \varepsilon_{\rm C} \ge \frac{4(1-q^2)\left(1+\sqrt{1+\frac{q^2}{2(1-q^2)}}\right)}{q^2} \right), \label{eq.copmaxj2_upper}
\end{cases}
\end{eqnarray}
\end{widetext}
where 
\begin{eqnarray}
R_4\equiv \sqrt{1+\frac{q^2}{2(1-q^2)}}.
\end{eqnarray}
We obtain the lower bound of $\varepsilon_{\rm maxJ_2}(q)$ in the asymmetric dissipation limit of 
$\gamma_h/\gamma_c \to \infty$ for any $|q|$-value:
\begin{eqnarray}
\varepsilon_{\rm maxJ_2}^-(q)=0.\label{eq.cop_maxJ2_lower}
\end{eqnarray}
These bounds are compared with those of the maximum COP in Fig.~\ref{lower_and_upper_bound_cop_max}.  
In this figure, we stress that as $q$ is decreased from unity, as is similar to the behavior of the heat engine in Fig.~\ref{gpc_p_eta_q}, the allowed range of the optimized COPs rapidly becomes smaller and the maximum COP and the COP at maximum cooling power regimes tend to collapse in a unique inefficient performance regime (smaller loop) as in Fig.~\ref{gpc_j2_cop_q}.

For the symmetric dissipation $\gamma_h=\gamma_c$, we obtain
\begin{equation}
\varepsilon_{\rm maxJ_2}^{\rm sym}(q)=\frac{q^2\varepsilon_{\rm C}-2(1-q^2)}
{2q^2 \left(1+\varepsilon_{\rm C}\right)}.
\end{equation}

Additionally, if the tight-coupling condition $|q|=1$ is fulfilled, we reproduce the results in~\cite{apertet13a}:
\begin{equation}\label{copmaxj2a}
\varepsilon_{\rm maxJ_2}^-=0 \,\,\left(\frac{\gamma_h}{\gamma_c}\rightarrow \infty, \ |q|=1 \right),
\end{equation}
\begin{equation}\label{copmaxj2b}
\varepsilon_{\rm maxJ_2}^{\rm sym}=\frac{\varepsilon_{\rm C}}{2(\varepsilon_{\rm C}+1)} \,\,\left(\frac{\gamma_h}{\gamma_c}\rightarrow 1, \ |q|=1 \right),
\end{equation}
\begin{equation}\label{copmaxj2c}
\varepsilon_{\rm maxJ_2}^+=\frac{\varepsilon_{\rm C}}{\varepsilon_{\rm C}+2} \,\,\left(\frac{\gamma_h}{\gamma_c}\rightarrow 0, \ |q|=1 \right).
\end{equation}
\section{Discussion and conclusions}
We have reported unified results for both HE and RE obtained by MNLIT models making emphasis on the influence of  irreversibilities by the heat leaks, internal dissipations, and their interplay. The results in Sec.~III clearly show that in order to obtain realistic (loop-shaped) power-efficiency and cooling power-COP performance characteristics, the irreversibilities by the heat leaks are a necessary ingredient. On the other hand, the internal dissipations into the heat reservoirs
account for quantitative variations in the involved energetic magnitudes but without affecting the qualitative behaviors.

In Sec.~IV, we have presented explicit calculations for some optimized figures of merit and their bounds in terms of dissipation to heat reservoirs and of the coupling parameter. 
In particular, the limiting results for $|q|=1$ deserve some comments. 
In this limit, the efficiency at maximum power Eq.~(\ref{remchig}) can be rewritten as
\begin{equation}
\eta_{\rm maxP}=\frac{\eta_{\rm C}}{2-\gamma \eta_{\rm C}},\label{eq.eta_ss}
\end{equation}
with $\gamma \equiv \frac{1}{1+\gamma_c/\gamma_h}=\frac{\gamma_h}{\gamma_c+\gamma_h}$, i.e., the ratio of the dissipation strength in the hot heat reservoir to the overall strength.
This result was also previously reported by Schmiedl and Seifert in a stochastic heat engine model~\cite{seifert08} and then reinterpreted
by Apertet et al.~\cite{apertet12b} as the characteristic efficiency at maximum power for exoreversible HE models where the only irreversibility comes from the internal dissipations.
This formula may also be connected to the LD models by interpreting the coefficients of dissipation-strength as the coefficients of heat-transfer between the working substance and the heat reservoir in 
the FTT framework~\cite{jchen13a}.  
Indeed, if we choose $\gamma=0$ and $1$ in Eq.~(\ref{eq.eta_ss}), we reproduce the bounds given by Eqs.~(\ref{remchilb}) and (\ref{remchiub}), respectively, while for symmetric dissipation $\gamma=1/2$ we reproduce
Eq.~(\ref{remchisyma}).
For RE, the COP at maximum cooling power Eq.~(\ref{eq.cop_maxJ2}) under $|q|=1$ is given as
\begin{eqnarray}
\varepsilon_{\rm maxJ_2}=\frac{\varepsilon_{\rm C}}
{2+\frac{\varepsilon_{\rm C}}{1-\gamma}},\label{eq.cop_mp}
\end{eqnarray}
by using $\gamma$. 
For $\gamma=0$ and $1$ we reproduce the bounds in Eqs.~(\ref{copmaxj2c}) and (\ref{copmaxj2a}), respectively, while for symmetric dissipation $\gamma=1/2$ we reproduce Eq.~(\ref{copmaxj2b}), 
which is already reported in~\cite{apertet13a} as a particular case of a thermoelectric refrigerator.
Therefore, the MNLIT model under the tight-coupling condition also reproduces correctly the results of this exoreversible model.

A special comment is merited by the maximum cooling power condition for RE expressed in Eq.~(\ref{eq.cop_mp}) for which Apertet et al.~\cite{apertet13a} have proposed that it should be considered as the only genuine counterpart of the Schmiedl-Seifert efficiency Eq.~(\ref{eq.eta_ss}) for HE. Both results are obtained under the same exoreversible conditions optimizing the efficiency (COP) under maximum useful benefit (power output for HE and cooling power for RE). On the other hand, the original CA value was obtained under quite different assumption of endoreversibility (without internal dissipations and heat leaks), which when reversed does not allow the optimization of the cooling power~\cite{apertet13a}.
Exoreversible and endoreversible models are indeed two (extreme) different models which define different specific coupling to the external heat reservoirs, thus it is not surprising that they lead to different expressions. Conversely, the LD models provide a unified framework for HE and RE where the exact CA-value emerges linked to a certain symmetric condition. In this context (with the same model, same symmetric condition, and same optimization criterion), 
$\varepsilon_{{\rm max\chi}}=\sqrt{1+\varepsilon_{\rm C}}-1\equiv \varepsilon_{\rm CA}$ for RE 
was proposed as the CA counterpart~\cite{carla12a,yuki13a}.
This value gives a better comparison with experimental results~\cite{carla13} than the COP at maximum cooling power Eq.~(\ref{eq.cop_mp}) whose maximum possible value is unity even in the limit of the small temperature difference as $\varepsilon_{\rm maxJ_2}\approx 1-\gamma\leq 1$ while the Carnot COP diverges as $\varepsilon_{\rm C}\to \infty$ in this limit.

In closing, all of the above clearly illustrates that the minimally nonlinear irreversible model succeeds in reproducing 
various results derived by previous studies.
In particular, the MNLIT model provides a clear interpretation of the global performance characteristics of generic heat devices in terms of the interplay between the heat leaks and the internal dissipations, and it reproduces the figures of merit optimized under some performance criteria and some conditions for both HE and RE.
Additionally, further studies are needed to establish clearer connections between the MNLIT models~\cite{yuki12a,yuki13a}, FTT frameworks~\cite{curzon75}, LD models~\cite{esposito10a}, and LIT models based on the local force-flux relationships~\cite{apertet13b}.
Related to this, we note that a recent work~\cite{sheng13a} reports a complementary idea of the nonlinear dissipation terms 
by introducing the concepts of weighted reciprocal temperatures and weighted thermal fluxes.
In~\cite{bizarro,B}, we also note that dissipation effects by the friction on the heat devices have been discussed. Although such friction effects as a cause of the dissipations into the heat reservoirs have not been taken into account in our present model, 
an extension of our model along this line would be interesting in terms of explaining behaviors of actual heat devices.

\begin{acknowledgements}
Y. I. acknowledges the financial support from a Grant-in-Aid for JSPS Fellows (Grant No. 25-9748). 
J. M. M. R. and A. C. H. acknowledge Ministerio de Economia y Competetividad of Spain, MINECO (ENE2013-40644-R).
\end{acknowledgements}

\appendix
\section{MNLIT formulation of LD model}
We consider the physical relevance of the nonlinear dissipation terms in the MNLIT model.
The MNLIT model was originally proposed with the motivation to explain and extend the low-dissipation Carnot cycle model~\cite{esposito10a} 
from a nonequilibrium thermodynamics point of view.
The low-dissipation Carnot cycle is a finite-time Carnot cycle model that runs a Carnot cycle at a finite rate, where the heat flowing during each isothermal process is given by~\cite{esposito10a}
\begin{eqnarray}
&&Q_h=T_h\Delta S-\frac{T_h\Sigma_h}{t_h},\label{eq.lowdissi1}\\
&&Q_c=T_c\Delta S+\frac{T_c\Sigma_c}{t_c},\label{eq.lowdissi2}
\end{eqnarray}
where $\Delta S$, $t_i$, and $\Sigma_i$ are the quasistatic entropy change of the working substance during the isothermal expansion process, the time duration, and the strength of the dissipation of each isothermal process, respectively. For simplicity, we here neglect the durations of the adiabatic processes~\cite{esposito10a}.
The entropy production rate of the present system $\dot{\sigma}=\frac{1}{t_h+t_c}\left(-\frac{Q_h}{T_h}+\frac{Q_c}{T_c}\right)$ is decomposed as follows:
\begin{eqnarray}
\dot{\sigma}&&=-\frac{1}{t_h+t_c}\frac{W}{T_c}+\dot{Q}_h\left(\frac{1}{T_c}-\frac{1}{T_h}\right)\nonumber\\
&&\equiv J_1X_1+J_2X_2,
\end{eqnarray}
where the thermodynamic fluxes and forces are defined as
\begin{eqnarray}
&&J_1\equiv \frac{1}{t_h+t_c}, \ X_1\equiv -\frac{W}{T_c},\label{eq.thermo_flux_force_lowdissi1}\\
&&J_2\equiv \dot{Q}_h, \ X_2\equiv \frac{1}{T_c}-\frac{1}{T_h}.\label{eq.thermo_flux_force_lowdissi2}
\end{eqnarray}
Under these definitions and using Eqs.~(\ref{eq.lowdissi1}) and (\ref{eq.lowdissi2}), we can easily show that $J_1$ and $J_2$ are transformed into the extended Onsager relations in the MNLIT model in Eqs.~(\ref{eq.exonsager_J1_he}) and (\ref{eq.exonsager_J2_he})~\cite{yuki12a}. 
Then the heat flux in each side of the isothermal process turns out to include the nonlinear dissipation term, 
which implies that the entropy production occurs equally in both sides of the isothermal processes.
The Onsager coefficients and the dissipation constants are given as~\cite{yuki12a}
\begin{eqnarray}
&&L_{ij}=
\begin{pmatrix}
\frac{T_c}{Y} & \frac{T_hT_c\Delta S}{Y} \\
\frac{T_hT_c\Delta S}{Y} & \frac{T_h^2T_c\Delta S^2}{Y}
\end{pmatrix},\\\label{eq.onsager_lowdissi}
&&\gamma_h=T_h\Sigma_h(\alpha+1), \ \gamma_c=\frac{T_c\Sigma_c(\alpha+1)}{\alpha},\label{eq.dissi_const}
\end{eqnarray}
where $Y\equiv (T_h\Sigma_h+T_c\Sigma_c/\alpha)(\alpha+1)$ and $\alpha \equiv \frac{t_c}{t_h}$.
From Eq.~(\ref{eq.onsager_lowdissi}), we can find that the Onsager reciprocity and the constraint in Eq.~(\ref{eq.restrict_onsager_coeffi_he}) hold even without
taking the limit of the small temperature difference $\Delta T\to 0$. The tight-coupling condition $|q|=1$ is also confirmed.
In the limit of $\Delta T\to 0$, we recover the ordinary linear Onsager relations in Eqs.~(\ref{eq.onsager_J1_he}) and (\ref{eq.onsager_J2_he}), where $J_2$ has no nonlinear dissipation term. This implies that our extended relations are a minimum extension of the Onsager relations.
The above consideration of the theoretical tight-coupling example led us to propose the more general MNLIT model including non-tight-coupling cases as the minimal model of nonlinear irreversible heat engines.
We can give such a nontight-coupling example described by the MNLIT model
as the following leaky low-dissipation Carnot cycle model (see Ref.~\cite{yuki13a} for its counterpart in refrigerators):
\begin{eqnarray}
&&Q_h=T_h\Delta S-\frac{T_h\Sigma_h}{t_h}+\kappa(T_h-T_c)(t_h+t_c),\\
&&Q_c=T_c\Delta S+\frac{T_c\Sigma_c}{t_c}+\kappa(T_h-T_c)(t_h+t_c),
\end{eqnarray}
where the last terms express the heat conduction between the two heat reservoirs according to the Fourier law with $\kappa$ being the thermal conductivity.
In this model, the thermodynamic fluxes and forces, and the dissipation constants are given by the same forms as in Eqs.~(\ref{eq.thermo_flux_force_lowdissi1}), (\ref{eq.thermo_flux_force_lowdissi2}) and (\ref{eq.dissi_const}), while the Onsager coefficients are modified as
\begin{eqnarray}
&&L_{ij}=
\begin{pmatrix}
\frac{T_c}{Y} & \frac{T_hT_c\Delta S}{Y} \\
\frac{T_hT_c\Delta S}{Y} & \frac{T_h^2T_c\Delta S^2}{Y}+T_hT_c\kappa
\end{pmatrix},\label{eq.onsager_matrix_2}
\end{eqnarray}
from which we obtain $|q|<1$.
This leaky low-dissipation Carnot cycle model indeed takes into account basic irreversibilities that exist in nonlinear irreversible heat engines: dissipation and heat leak.
These irreversibilities are also taken into account by steady state irreversible heat engines such as thermoelectric devices~\cite{apertet12b}.
This implies that the cyclic heat engines and steady-state heat engines 
with the above basic irreversibilities are unified in terms of our MNLIT model. Completely the same arguments can be applied to refrigerators.


\begin{references}
\bibitem{monotonicity} We have confirmed the monotonicity numerically, though we have not analytically.
\end{references}
\end{document}